\documentclass[preprint,12pt]{elsarticle}
\usepackage{color}

\usepackage{amssymb}
\journal{Solid state commun.}

\begin{document}

\begin{frontmatter}



\title{Study of two-spin entanglement in singlet states}


\author{M. Q. Lone, A. Dey}
\author{S. Yarlagadda\corref{*}}
 \ead{y.sudhakar@saha.ac.in}
 \cortext[*]{Corresponding author}
\address{$CMP$ Division, Saha Institute of Nuclear physics\\
1/AF Salt Lake, Kolkata 700064, India}

\begin{abstract}
We study the entanglement properties of 
two-spin subsystems
 in spin-singlet states.
The average entanglement between two spins is maximized in a single valence-bond (VB) state.
On the other hand, 
 $E_v^2$ (the average  entanglement between a subsystem
of two spins and the rest of the system)  
can be maximized through a homogenized superposition of the VB  states.
The maximal $E_v^2$ rapidly increases with system size and 
saturates at its maximum allowed value. 
 We adopt two ways of obtaining maximal $E_v^2$ states:
(1) imposing homogeneity on singlet states; 
 and (2) generating isotropy in a general homogeneous state.
 By using these two approaches, we construct explicitly  four-spin and six-spin highly 
entangled states that are both  isotropic and homogeneous. 
Our maximal  $E^2_v$ states represent a new class of 
resonating-valence-bond states which we show to be the ground states of 
the infinite-range
%, isotropic
 Heisenberg model.
\end{abstract}

\begin{keyword}
D. Singlet states \sep D. RVB states \sep D. Entanglement \sep A. Frustrated magnets 
%\PACS 75.10.Kt \sep 75.10.Jm \sep  03.67.Mn \sep 03.67.Bg



\end{keyword}

\end{frontmatter}


\section{Introduction}
%\label{}
Valence-bond states 
were shown
to be the ground states of spin systems
earlier by Majumdar's group \cite{ckm} and later by Shastry and Sutherland \cite{shastry}.
Any 
spin-singlet state (i.e., state with total spin eigenvalue $S_T=0$)
  can be expressed
 as a superposition of VB states {\cite{hulthen,suppl}}.
Spin-singlet 
 states are important in understanding many problems in  condensed matter physics
as well as in quantum information science.
Resonating-valence-bond (RVB) singlet states 
have provided interesting insights for understanding strongly correlated phenomena 
such as spin liquid physics in frustrated magnets \cite{fazekas},
 physics of high $T_c$ cuprates \cite{anderson,baskaran,lee},
superconductivity in organic solids \cite{ishiguro}, insulator superconductor 
transition in boron-doped diamond \cite{ekimov}, etc. Furthermore, RVB states have also
been proposed as robust basis states for topological quantum computation \cite{kitaev}.
{As regards examples of real systems, valence-bond state (comprised of alternating dimer spin chain) has been experimentally realized in the
spin-chain compound, copper
nitrate (${\rm Cu(NO_3)_2 \times  2.5 H_2 O}$) \cite{chiranjib}; RVB state occurs naturally
in ${\rm SrCu_2(BO_3)_2}$  \cite{ueda} which can be explained by the Shastry-Sutherland model \cite{shastry2}; 
valence bond solid has been observed in ${\rm Zn_x Cu_{4-x} (OD)_6 Cl_2}$ \cite{sachdev}.}

Correlation functions of observables
(such as density, magnetization, etc.) 
reflect  the degree of entanglement in the pure state of a many-body system 
\cite{latorre}.
Quantum algorithms that would significantly accelerate  a classical computation
must rely on highly entangled states since  slightly entangled states
can be simulated efficiently on a classical computer \cite{vidal}.
Thus, characterization of multi-particle entanglement and production
of 
maximal
 entanglement is vital for 
quantum computational studies and for mutual enrichment of
 quantum information science and many-body condensed matter physics.

Correlation/entanglement between two spins plays an important role in understanding
 phase transitions, length scale in the system, etc.
Although two-spin correlation/entanglement has  been investigated in certain 
RVB states \cite{LA,sen},
%that were proposed as states close in energy to the ground state
% of the Heisenberg Hamiltonian \cite{LA,sen},
 to our knowledge,
 there has been no explicit construction of  RVB states that would  contain
maximal entanglement of two-spin subsystems. 
The spins of a two-spin singlet, while being maximally entangled with each other, are completely unentangled
with the remaining spins and thus show monogamy. Thus, if we wish to establish greater
entanglement between the two-spin subsystem  and the rest of the spin system,
we are forced to diminish entanglement between the  spins of the two-spin subsystem.
The purpose of the present paper is to 
enhance our understanding of
the distribution of two-spin entanglement in singlet states.
We analyze the following two extreme cases in a general singlet: 
(1) maximal average entanglement between two spins;
and (2) maximal average entanglement between
a two-spin subsystem and the remaining spins.
 The main results of this paper are as follows. First, we study  two-spin entanglement in singlets.
 We show that  the average entanglement between two spins is maximum (as expected) for a single VB state. 
In a singlet, we also demonstrate that
  ${\rm SU(2)}$ isotropy and  homogeneity (in spin-spin correlation function) maximize the bipartite
 entanglement $E_v^2$ while minimizing the average entanglement between two spins.
 Second, based on the principles of isotropy and homogeneity, we propose two approaches to 
construct  entangled states that maximize $E_v^2$
and are a new class of RVB states.
 Last, we analyze these states in terms of ground states of
the infinite-range Heisenberg model (IRHM).




\section{Entanglement for two spins calculated from reduced density matrix}
% of spin singlet}
For a bipartite system AB in a pure state,
von Neumann entropy $E_v$ measures the entanglement between the subsystems A
and B.
From the reduced density matrices 
 $\rho_A \equiv {\rm Tr}_B {\rho^{AB} }$ and $ \rho_B \equiv {\rm Tr}_A {\rho^{AB} }$,
obtained from the pure state $\rho^{AB}$, we get the following:
\begin{eqnarray}
\!\!\!\! E_v(\rho_{A}) &=& -{\rm Tr}(\rho_A \log_2 \rho_A)
\nonumber \\ 
                       &=& -{\rm Tr}(\rho_B \log_2 \rho_B) = E_v(\rho_B).
\end{eqnarray}
Using $ S^i=\frac{1}{2}\sigma^i $ and the basis $|\downarrow\rangle$ and $|\uparrow\rangle$, 
the single 
spin (reduced) density matrix can be written as \cite{21}
\begin{eqnarray}
\rho_i= \left[
\begin{array}{cc}
\frac{1}{2}-\langle S^z_i \rangle & \langle S^{+}_i \rangle\\

\langle S^{-}_i \rangle  & \frac{1}{2}+\langle S^z_i \rangle\\
\end{array}
\right] ,
\label{rhoi}
\end{eqnarray}
where $S^\pm \equiv S^x \pm i S^y$.
For ${\rm SU(2)}$ singlet states, since
the z-component of the total spin operator ($S^z_{Total}$) has  eigenvalue $S^z_T=0$,
$\langle S^{+}_i \rangle = 0$.
Then, from the isotropy of the singlet states,
it follows that 
$\langle S^z_i \rangle =0$. Consequently,
the single-spin density
 matrix  becomes maximally mixed and the  entanglement $E_v(\rho_i)$ becomes maximized.

We will now show that the average entanglement between a pair of spins is  maximum
for a VB state.
To this end, we obtain an expression for the tangle which is a measure of entanglement. 
For a pure state, the tangle between the spin at $i$ and the rest of the spins 
 is given
by $\tau(\rho_i) = 2 [1- {\rm Tr}(\rho_i^2)]$ \cite{verstraete}; from Eq. (\ref{rhoi}),
it follows that $\tau(\rho_i) =1$ for a spin singlet.
%On the other hand,  for the mixed state $\rho_{ij}$ 
%(i.e., the reduced density matrix for spins at $i$ and $j$), the tangle $\tau(\rho_{ij})$ 
Next, we will deal with the tangle between spins at $i$ and $j$
for the mixed state $\rho_{ij}$ (i.e., the two-spin reduced density matrix).
The tangle $\tau(\rho_{ij})$ 
is given by the square of ${\rm max}\{\lambda_1-\lambda_2-\lambda_3-\lambda_4,0\}$ where
$\lambda_1$, $\lambda_2$, $\lambda_3$, and $\lambda_4$ are the square roots of the eigenvalues
(in decreasing order) of the operator 
$\rho_{ij} (\sigma_y \otimes \sigma_y) \rho^{\star}_{ij} (\sigma_y \otimes \sigma_y)$
where the asterisk corresponds to complex conjugation in the basis 
$\{ |00\rangle, ~|01\rangle,~|10\rangle,~|11\rangle\}$. 
Now, the distribution of the bipartite entanglement (as measured by the tangle) amongst N spins
satisfies the following inequality \cite{verstraete}: 
\begin{eqnarray}
\sum_{j \neq i} \tau(\rho_{ij}) \le \tau(\rho_i) . 
\end{eqnarray}
For a VB state,
it is straight forward to show
 that
 the tangle $\tau (\rho_{ij})=1$ 
 when the spins at $i$ and $j$ form a singlet
and that  $\tau (\rho_{ij})=0$ otherwise. Thus, we see that a VB state satisfies 
$ \sum_{j \neq i} \tau(\rho_{ij}) = \tau(\rho_i)$ for all $i$. Hence,
in a VB state, the average entanglement between two spins $ [1/N(N-1)]\sum_{i,j \neq i} \tau(\rho_{ij})$ 
attains the maximum allowed valued of 1.

\begin{figure}
%[t]
\begin{center}
\includegraphics[angle=90,angle=90,angle=90,width=3.0in,height=2.0in]{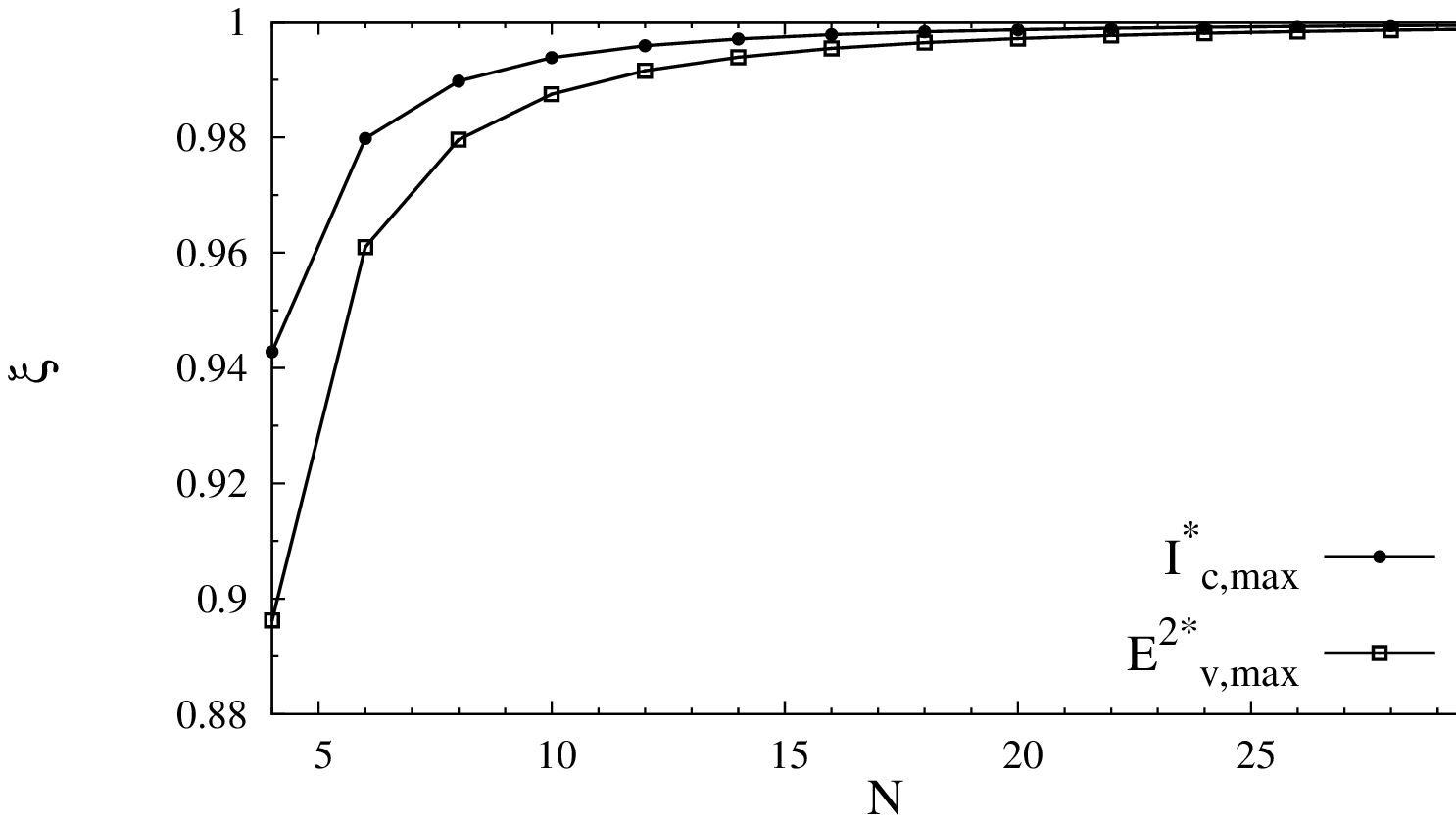}
\caption{Normalized entanglement 
$\xi$, for the two-spin reduced density matrix, measured
by 
 (a) von Neumann entropy [${\rm E^{2 \star}_{ v,max}} \equiv {\rm E^2_{ v,max}(N)/E^2_{v, max}}(\infty)$]
 and
 (b) i-concurrence [${\rm I_{c, max}^{\star}} \equiv {\rm I_{c, max}(N)/I_{c, max}(\infty)}$], 
for
 N-spin
 VB systems. }
\label{e_v}
\end{center}
\end{figure}

We will now proceed to obtain the condition for maximal average entanglement
between a two-spin subsystem and the rest of the spins in
a singlet.
On realizing that $\langle S^z_i \rangle =0$, we obtain the following expression for the 
two-spin reduced density matrix \cite{21}:
\begin{eqnarray}
\!\!\!\!\!\!\!\!\!\!\!\! \rho_{ij} \!= \!\left[\!
\begin{array}{cccc}
\frac{1}{4} + \langle S^z_i S^z_j \rangle  & 0 & 0 & 0\\
0 & \!\frac{1}{4} - \langle S^z_i S^z_j \rangle \! & \langle S^+_i S^-_j\rangle   & 0\\
0 &  \langle S^-_i S^+_j\rangle  & \!\frac{1}{4}- \langle S^z_i S^z_j \rangle \!& 0\\
0 & 0 & 0 & \!\frac{1}{4} + \langle S^z_i S^z_j \rangle 
\end{array}
\!\right] . 
\nonumber \\
\label{d_mat}
\end{eqnarray}
Here, isotropy implies that
$0.5 \langle S^-_i S^+_j \rangle = 0.5 \langle S^+_i S^-_j \rangle 
=\langle S^x_i S^x_j \rangle = \langle S^y_i S^y_j \rangle 
= \langle S^z_i S^z_j \rangle $.
Thus, the von Neumann entropy 
$ E_v(\rho_{ij}) $
 can be expressed as
\begin{eqnarray}\label{Eij}
\!\!\!\!\!\!\! E_v(\rho_{ij}) = 2 - \frac{1}{4} \!\!\!&[& \!\!\!3 (1+
4 \langle S^z_i S^z_j \rangle 
)\log_2(1+
4 \langle S^z_i S^z_j \rangle 
) 
\nonumber \\
&& \!\!\!\!\! + (1-
12 \langle S^z_i S^z_j \rangle 
 ) \log_2(1-
12 \langle S^z_i S^z_j \rangle 
  )
] .                              
\end{eqnarray} 
For our states, since $S^z_{Total}|\Psi \rangle = 0$, we observe that
% $\langle S^{z}_i \sum_j S_j^{z}\rangle=0$, i.e.,
\begin{eqnarray}
\sum_{j \neq i}\langle S^{z}_{i} S^{z}_{j}\rangle=-\langle {S^{z}_i}^2 \rangle=-\frac{1}{4} .
\label{Sz_const}
\end{eqnarray}
We will now maximize the total entanglement entropy
$\sum_{i,j \neq i} E_v(\rho_{ij})$ subject to the  constraint in Eq. (\ref{Sz_const}).
To this end, we will employ the method of Lagrange multipliers and define the Lagrange
function $\Lambda$ as follows:
\begin{eqnarray}
\Lambda = \sum_{i,j \neq i} E_v(\rho_{ij}) - \sum_{i} 
\lambda_i \left ( \sum_{j \neq i}\langle S^{z}_{i} S^{z}_{j}\rangle+\frac{1}{4} \right ) .
\label{Lambda}
\end{eqnarray}
Then, setting $\frac{\partial \Lambda}{\partial \langle S^{z}_l S_m^{z}\rangle} =0$
yields 
\begin{eqnarray}
\lambda_l = 3 \log_2 \left [\frac{(1-12 \langle S^{z}_l S_m^{z}\rangle)}{(1+4  \langle S^{z}_l S_m^{z}\rangle)}
\right ] ,
\label{lambda_i}
\end{eqnarray}
which implies that the optimal $\langle S^{z}_{l} S^{z}_{m}\rangle $ is independent of $m$
for all $m \neq l$.
Consequently, 
 it follows from Eq. (\ref{Sz_const}) that $\sum_{i,j \neq i} E_v(\rho_{ij})$ is maximized
when $\langle S^{z}_{i} S^{z}_{j}\rangle=-\frac{1}{4(N-1)}$,
i.e., when the isotropic state has a homogeneous longitudinal spin-spin correlation function.   
The
average entanglement entropy 
between the subsystem of two spins 
 and the remaining
$N-2$ spins, expressed as $E^2_v \equiv [1/N(N-1)] \sum_{i,j \neq i} E_v(\rho_{ij}) $,
has a maximum value given by
\begin{eqnarray}\label{E_vN}
\!\!\!\! E_{v,max}^2(N) &=&   -3\left( \frac{1}{4} - \frac{1}{4(N-1)} \right)
\log_2\left( \frac{1}{4} - \frac{1}{4(N-1)} \right)  \nonumber\\
&&-\left(\frac{1}{4} + \frac{3}{4(N-1)} \right)
\log_2\left(\frac{1}{4} +\frac{3}{4(N-1)} \right) . 
\nonumber \\
\end{eqnarray}
It is interesting to note that for $N \rightarrow \infty$, 
the above expression yields $E_v^2 \rightarrow 2$.
In fact, $E_v^2$ approaches the maximum possible value $2$ quite rapidly as can be seen
from Fig.~\ref{e_v}.
Next, for $N=4$, we observe that our expression for $E_{v,max}^2$ in Eq. (\ref{E_vN})
yields the same entanglement
entropy value $1+0.5 \log_2 3$ as that obtained for the four-spin maximally entangled
 state studied in Refs. \cite{HS,plastino}.
%Furthermore, our approach (showing that $E^2_v$ is maximized for homogeneous 
%spin-spin correlation function)
%explains why all the pairs of this state give the same entanglement value.
%Contrastingly, for the isotropic ground state,
%in the  case of a Heisenberg chain with nearest-neighbor
% interaction, for four spins
% the entanglement  entropy $E^2_v =1.21$ (with $\langle S^z_i S^z_j \rangle= -0.5/3 $)
% while for an infinite chain  the von Neumann entropy $E^2_v = 1.37$ 
%(with $\langle S^z_i S^z_j \rangle \approx -0.443/3 $) 
%both of which  are far less than our maximal $E^2_v$ values above. 
%It is of interest to note that, when
%$N=4$ or $N \rightarrow \infty$, the maximal values of $E^2_v$ for isotropic systems
%are the same as the maximal $E^2_v$ values for general systems (i.e., systems that
%can be either isotropic or non-isotropic).

We also note that homogeneity in the two-spin correlation function $\langle S^{z}_{i} S^{z}_{j}\rangle$,
 under the constraint of Eq. (\ref{Sz_const}),
 maximizes the average entanglement (between a two-spin subsystem and the rest of the system)
as measured by i-concurrence ${\rm I_c}$ given below  \cite{rungta}: 
\begin{eqnarray}
 {\rm I_c} = \frac{2}{N(N-1)}\sum_{i,j> i} \sqrt{2[1-{\rm Tr}(\rho_{ij}^2)]} .
\end{eqnarray}
As shown in Fig. 1, ${\rm I_{c, max}}$ also monotonically increases with system size.

Lastly, we mention that the tangle $\tau(\rho_{ij})$ (between
spins at $i$ and $j$) for a singlet
[on using the definition
and Eq. (\ref{d_mat})] is given by square of 
${\rm max}\{4|\langle S^{z}_{i} S^{z}_{j}\rangle|-2 \langle S^{z}_{i} S^{z}_{j}\rangle-1/2,0\}$.
For the singlets that have $E^2_v$ maximized 
(which occurs when $\langle S^{z}_{i} S^{z}_{j}\rangle=-\frac{1}{4(N-1)}$),  
 the tangle $\tau(\rho_{ij})=0$ when $N \ge 4$.
Hence, the average entanglement between spins at sites $i$ and $j$
is given by $ [1/N(N-1)]\sum_{i,j \neq i} \tau(\rho_{ij})=0$.

Our maximal $E^2_v$ states can be regarded as a new class of RVB states made of homogenized
superposition of 
%isotropic $S_T =0$
 valence-bond states.
In the related work of Ref. \cite{sen}, an interesting analysis
of entanglement between two sites was carried out in two different
RVB systems, i.e., the RVB gas 
involving equal-amplitude superposition of all bipartite VB coverings and the RVB liquid
involving equal-amplitude superposition of all  nearest-neighbor-singlet
 VB coverings of a lattice.
The RVB gas, 
%(involving equal amplitude superposition of all bipartite VB coverings)
for  finite-size systems, revealed
 a non-zero tangle between  two sites. 
On the other hand, the RVB liquid, 
%(involving equal amplitude superposition of all  nearest-neighbor
%singlet VB coverings of a lattice)  
for small systems, i.e., a $4 \times 4$ lattice, manifests 
zero (non-zero) tangle
between  two sites for open (periodic) boundary conditions \cite{suppl}.

\section{Maximal $ E^2_v $  states}
In this section, we offer two different
approaches for constructing maximal 
$E_v^2$ states. The first approach involves producing 
homogeneity in isotropic states while the second deals with
generating isotropy in homogeneous states. As 
examples of our prescribed procedure,
we construct highly entangled states for four spins and six spins.

\subsection{Generating  homogeneity in isotropic states}

In this section, we will proceed to construct 
entangled states for N-spins that maximize $E_v^2$.
We first note that there are  $(N-1)!!$ states with $S_T=0$
and each of these is a product of $N/2$ 
two-spin singlet states of the form
{\nolinebreak$|\uparrow\downarrow\rangle-|\downarrow\uparrow\rangle$}.
% (with no pair of two-spin singlets sharing a spin). 
Of these
$(N-1)!!$ 
 product combinations with $S_T =0$,
  only ${^N}C_{\frac{N}{2}}-{^N}C_{\frac{N}{2} - 1} = N!/[(N/2)!(N/2+1)!]$ 
products
are linearly independent \cite{pauling}. A particular set of
 linearly-independent $S_T=0$ states are the Rumer states 
 that are made up of non-crossing singlets \cite{pauling,rumer}.

Next, we show how to construct highly entangled states 
by starting
with isotropic states and making them homogeneous. 
Since we are dealing with spin-singlets, we consider
$\frac{N}{2}$ $\uparrow$ spins and $\frac{N}{2}$ $\downarrow$ spins
in our basis states $ |\psi^z_k\rangle$ ($\equiv |\sigma^z_1 \sigma^z_2...\sigma^z_N \rangle$ with
$\sigma^z_i = \uparrow$ or $\downarrow$). Using superposition of these  
basis states, with  all the basis states being equally probable, we construct the general homogeneous states
\begin{eqnarray}
|\Psi_N\rangle_{hom}= \sum_{k=1}^{^N C_{N/2}} e^{i \delta_k}|\psi^z_k\rangle ,
\label{hom}
\end{eqnarray}
which we will now prove to yield $ \langle S^z_i S^z_j\rangle= - \frac{1}{4(N-1)}$. 
 Throughout this paper, for convenience, we ignore the normalization
constants in our spin states. 
Let the spin at site $i$ be either $\uparrow$ or $\downarrow$. Then,
 the probability 
that the spin at site $j \neq i$ is in the  same state is $\frac{[(N/2) - 1]}{[(N/2)-1]+N/2}$, whereas
% while 
the probability that it is in the  opposite state is $\frac{[N/2]}{[(N/2)-1]+N/2}$.
 Therefore,  we can write
\begin{eqnarray}
\langle S^z_i S^z_j\rangle &=& \frac{1}{4}\left[
 \frac{\frac{N}{2} -1}{(\frac{N}{2} -1) + \frac{N}{2}} 
\right] 
- \frac{1}{4}
\left[ 
 \frac{\frac{N}{2} }{(\frac{N}{2} -1) + \frac{N}{2}}
\right] \nonumber \\
&=& - \frac{1}{4(N-1)}.
\end{eqnarray}
Thus, when  all the basis 
states are equally probable, we get 
$ \langle S^z_i S^z_j\rangle= - \frac{1}{4(N-1)}$. 
Now, we are in a position  to form (from isotropic $S_T =0$ states)
homogenized states that are highly entangled.

\subsubsection{Maximal $E^2_v$ states for four spins}
\begin{figure}[]
\begin{center}
\includegraphics[width=2.0in,height=1.0 in]{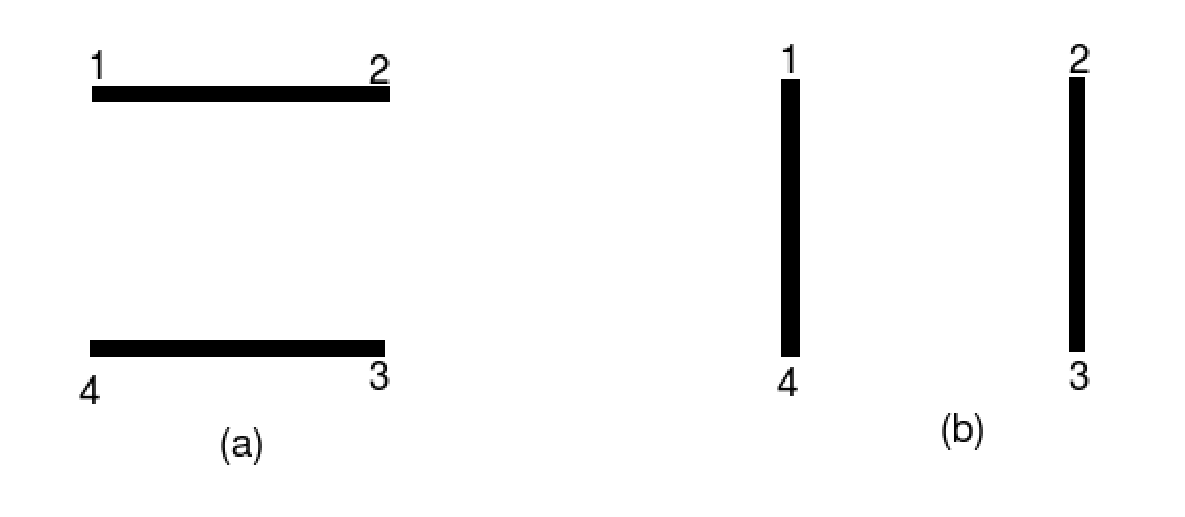}
\caption{Linearly-independent, non-crossing Rumer diagrams for a four-spin system . }
\label{four}
\end{center}
\end{figure}

For $N=4$ spins, we have two linearly-independent, non-crossing Rumer diagrams    which are shown in 
Fig. \ref{four}. The states corresponding to these
diagrams can be expressed as 
$|\Phi^{S_{12} =0}_{12}\rangle \otimes |\Phi^{S_{34} =0}_{34}\rangle$ and
$ |\Phi^{S_{14} =0}_{14}\rangle \otimes |\Phi^{S_{23} =0}_{23}\rangle$
where   $|\Phi^{S_{ij} =0}_{ij}\rangle \equiv  [ |\uparrow \rangle_i|\downarrow\rangle_j
- |\downarrow\rangle_i|\uparrow\rangle_j]  $ is a two-spin singlet state
% for spins at sites $i$ and $j$ 
with $S_{ij}$ being the total spin of $S_i$ and $S_j$.
It is worth noting that
\begin{eqnarray}
|\Phi^{S_{13} =0}_{13}\rangle \otimes |\Phi^{S_{24} =0}_{24}\rangle &=&
 |\Phi^{S_{12} =0}_{12}\rangle \otimes |\Phi^{S_{34} =0}_{34}\rangle
\nonumber \\
&& +
 |\Phi^{S_{14} =0}_{14}\rangle \otimes |\Phi^{S_{23} =0}_{23}\rangle  ,
\label{lin_dep}
\end{eqnarray}
which means that the crossing singlet $|\Phi^{S_{13} =0}_{13}\rangle \otimes |\Phi^{S_{24} =0}_{24}\rangle$
 depends linearly on the
two non-crossing singlets $|\Phi^{S_{12} =0}_{12}\rangle \otimes |\Phi^{S_{34} =0}_{34}\rangle$
and $|\Phi^{S_{14} =0}_{14}\rangle \otimes |\Phi^{S_{23} =0}_{23}\rangle $.
Using the above relation, one can establish that there are
only ${^N}C_{\frac{N}{2}}-{^N}C_{\frac{N}{2} - 1}$ linearly-independent
$S_T =0$ states.

\begin{table}[]
\begin{center}
    \begin{tabular}{ | l | l |}
    \hline
     Basis states  $|\psi^z_k\rangle$  &    Coefficients $e^{i\delta_k}$ in Eq. (\ref{hom}). \\ [1ex] \hline
    $\mid \uparrow\downarrow\uparrow\downarrow\rangle$ & 
         $ r_1 e^{i \theta_1} - r_2 e^{i \theta_2}$  \\ [1ex]\hline
    $\mid \downarrow\uparrow\downarrow\uparrow\rangle $ &  $r_1 e^{i \theta_1} - r_2 e^{i \theta_2}$ 
   \\  [1ex]     \hline
    $\mid \uparrow\downarrow\downarrow\uparrow \rangle $ & $-r_1 e^{i \theta_1}$ \\ [1ex]\hline
   $\mid  \downarrow\uparrow\uparrow\downarrow \rangle $  & $-r_1 e^{i \theta_1}$ \\ [1ex] \hline
   $\mid \uparrow\uparrow\downarrow\downarrow \rangle $ & $r_2 e^{i \theta_2 }$  \\ [1ex]\hline
   $\mid \downarrow\downarrow\uparrow\uparrow \rangle$ & $r_2 e^{i \theta_2}$   \\ [1ex]\hline
    \end{tabular}
\end{center}
\caption{Basis states $|\psi^z_k\rangle$ of Eq. (\ref{hom}) for a
four-spin system and the corresponding  coefficients $e^{i\delta_k}$
obtained by setting $|\Psi^4\rangle$ [in Eq. (\ref{fh})] equal to $ |\Psi_N \rangle_{hom}$ 
[in Eq. (\ref{hom})] with $N=4$.}
\label{table1}
\end{table}

Next, we take the following linear superposition 
 to obtain
the desired entangled state:
\begin{eqnarray}
|\Psi^4\rangle &=& r_1 e^{i \theta_1} \left[
 |\Phi^{S_{12} =0}_{12}\rangle \otimes |\Phi^{S_{34} =0}_{34}\rangle\right]  \nonumber \\
 &+& r_2 e^{i \theta_2 } \left[|\Phi^{S_{14} =0}_{14}\rangle \otimes |\Phi^{S_{23} =0}_{23}\rangle \right],
\label{fh}
\end{eqnarray}
where $r_i e^{i \theta_i}$ represents a general coefficient. 
%Evaluating  Eq. (\ref{fh}) (and 
On setting $|\Psi^4\rangle = |\Psi_N \rangle_{hom}$ with $N=4$,  
as shown in Table \ref{table1},
we get the various  coefficients
% (shown in Table \ref{table1})
 for each of the basis states $|\psi^z_k\rangle$ 
occurring  
in Eq. (\ref{hom}).

The expression for $|\Psi^4 \rangle$ [in Eq. (\ref{fh})] assumes
the homogenized superposed form shown in  
Eq. (\ref{hom}) for two linearly-independent sets of solutions of the
 coefficients in Table \ref{table1}. The  two sets of 
solutions are $\{r_1e^{i \theta_1}=- e^{i 2\pi/3}, r_2 e^{i\theta_2}= e^{i 4\pi /3}\}$
and its complex conjugate. The
states corresponding to these solutions are 
\begin{eqnarray}
|\Psi^4\rangle&=&    e^{i 2\pi/3} (|\Phi^{S_{12} =0}_{12}\rangle
 \otimes |\Phi^{S_{34} =0}_{34}\rangle)  \nonumber \\
&+&e^{i 4\pi /3}(|\Phi^{S_{14} =0}_{14}\rangle
 \otimes |\Phi^{S_{23} =0}_{23}\rangle) 
\nonumber \\ 
  &=& 
\left [
 (|\uparrow\downarrow\uparrow\downarrow \rangle +
 |\downarrow\uparrow\downarrow\uparrow \rangle) 
+ \omega_3(|\uparrow\downarrow\downarrow\uparrow\rangle +
|\downarrow\uparrow\uparrow\downarrow\rangle ) 
\right .
\nonumber\\
&&~~~~~~~~
+ \omega_3^2 
\left .
 (|\uparrow\uparrow\downarrow\downarrow \rangle
 +|\downarrow\downarrow\uparrow\uparrow \rangle) \right ] ,
\label{psi_hs}
\end{eqnarray}
and its complex conjugate. 
In the above equation,  $\omega_3=e^{i 2\pi /3}$, i.e., a  cube root of unity.
The above state is  the 
 maximally entangled state  for four spins studied in Refs. \cite{HS,plastino}.

\subsubsection{Maximal $E^2_v$ states for six spins} 
\begin{figure}[]
\begin{center}
\includegraphics[width=2.5in,height=2.0in]{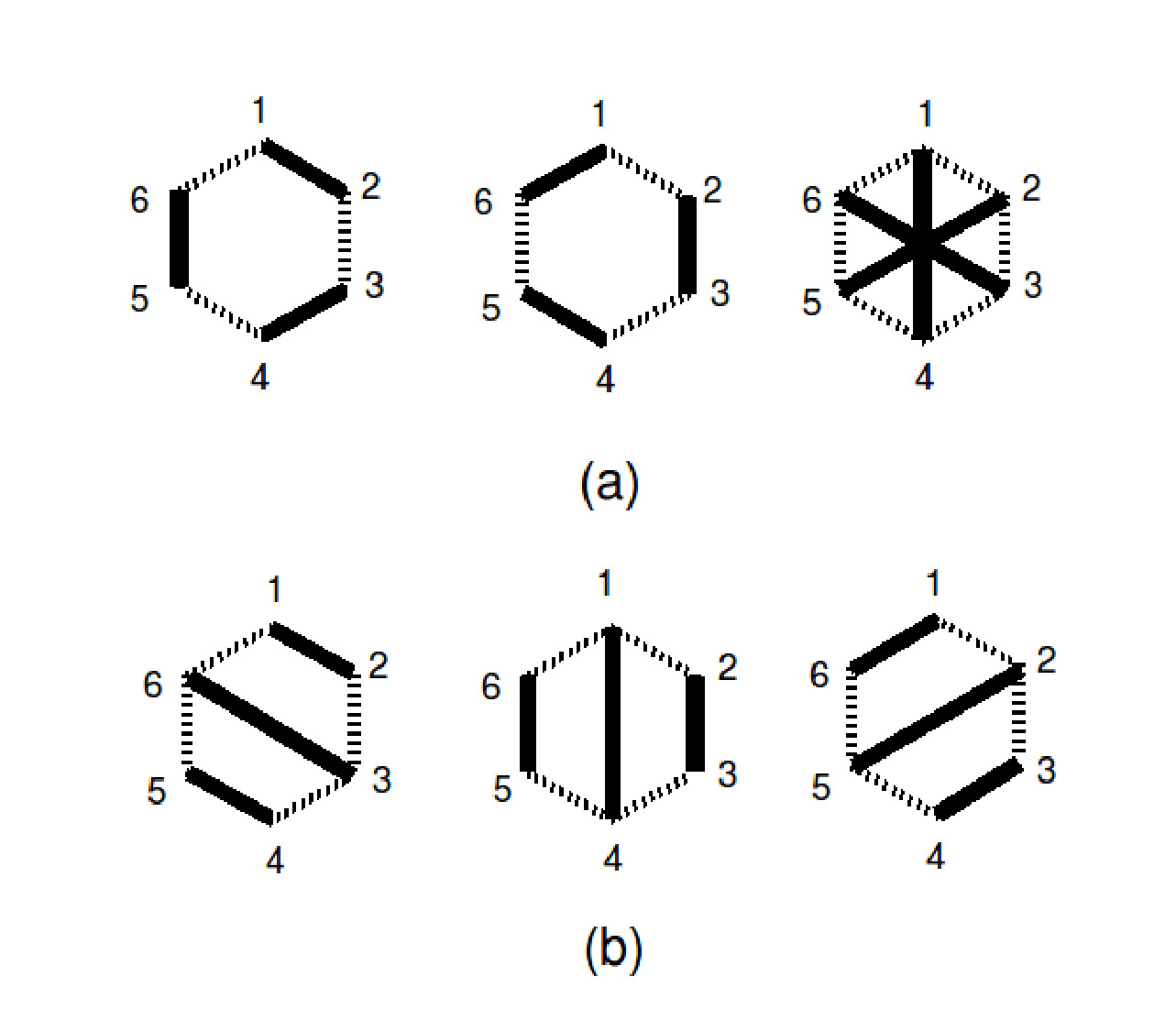}
\caption{Homogenized linear combination of the five non-crossing $S_T=0$ singlet states (or non-crossing Rumer diagrams) in 
 (a) and (b) give  
 maximal $E^2_v$ entanglement for six spins. The state $|\Psi_c^6\rangle$ in Eq. (\ref{rsol4}) is a linear combination
of the three diagrams in (a). On the other hand, $|\Psi_a^6\rangle$ in Eq. (\ref{sol1}) 
and $|\Psi_b^6\rangle$ in Eq. (\ref{sol3})
%, respectively,
  represent different linear combinations of all the three diagrams in (b).}
\label{six}
\end{center}
\end{figure}

The linearly-independent 
states in this case 
are five in number (i.e., the five non-crossing Rumer diagrams of
 Fig. \ref{six}).
 Thus, similar to the four-spin case, we begin with 
the following general linear superposition:
\begin{eqnarray}
|\Psi^6\rangle &=& r_1 e^{i \theta_1} \left[ 
| \Phi^{S_{12} =0}_{12} \rangle \otimes
 |\Phi^{S_{34} =0}_{34} \rangle \otimes
 |\Phi^{S_{56} =0}_{56}\rangle \right]\nonumber \\
&+& r_2 e^{i \theta_2} \left[
|\Phi^{S_{61} =0}_{61}\rangle \otimes
 |\Phi^{S_{23} =0}_{23} \rangle \otimes
 |\Phi^{S_{45} =0}_{45}\rangle \right] \nonumber \\
&+& r_3 e^{i \theta_3} \left[
 |\Phi^{S_{12} =0}_{12}\rangle \otimes |\Phi^{S_{36} =0}_{36} \rangle \otimes |\Phi^{S_{45} =0}_{45}\rangle \right] \nonumber \\
&+& r_4 e^{i \theta_4} \left[
| \Phi^{S_{23} =0}_{23} \rangle \otimes |\Phi^{S_{14} =0}_{14} \rangle \otimes |\Phi^{S_{56} =0}_{56}\rangle \right] \nonumber \\
&+&r_5 e^{i \theta_5} \left[
|\Phi^{S_{16} =0}_{16} \rangle \otimes |\Phi^{S_{25} =0}_{25}\rangle \otimes |\Phi^{S_{34} =0}_{34}\rangle \right], 
\label{sh}
\end{eqnarray}
and find the solutions that homogenize $|\Psi^6\rangle$, i.e., make $|\Psi^6\rangle$
assume the form of $|\Psi_6\rangle_{hom}$ in Eq. (\ref{hom}). 
Since there are 
only five linearly-independent Rumer states, we observe that we can construct at most five linearly-independent 
superpositions of the form 
of $|\Psi^6\rangle$
 in Eq. (\ref{sh}).
A set of five  linearly-independent solutions are as follows (with details given in Ref. \cite{suppl}):
\begin{eqnarray}
 |\Psi^6_a\rangle &=& \omega_4
 (|\Phi^{S_{12} =0}_{12}\rangle \otimes |\Phi^{S_{36} =0}_{36} \rangle \otimes |\Phi^{S_{45} =0}_{45}\rangle) \nonumber \\
&& + \omega_4^2(| \Phi^{S_{23} =0}_{23} \rangle \otimes |\Phi^{S_{14} =0}_{14} \rangle \otimes |\Phi^{S_{56} =0}_{56}\rangle) \nonumber \\
&& + \omega_4^3(|\Phi^{S_{16} =0}_{16} \rangle \otimes |\Phi^{S_{25} =0}_{25}\rangle \otimes |\Phi^{S_{34} =0}_{34}\rangle) ,
\label{sol1}
\end{eqnarray}
and its complex conjugate $ |\Psi^{6*}_a\rangle$,
\begin{eqnarray}
 |\Psi^6_b\rangle &=& - (|\Phi^{S_{12} =0}_{12}\rangle \otimes |\Phi^{S_{36} =0}_{36} \rangle \otimes |\Phi^{S_{45} =0}_{45}\rangle) \nonumber \\
&& + (| \Phi^{S_{23} =0}_{23} \rangle \otimes |\Phi^{S_{14} =0}_{14} \rangle \otimes |\Phi^{S_{56} =0}_{56}\rangle) \nonumber \\
&& - (|\Phi^{S_{16} =0}_{16} \rangle \otimes |\Phi^{S_{25} =0}_{25}\rangle \otimes |\Phi^{S_{34} =0}_{34}\rangle) ,
\label{sol3}
\end{eqnarray}
\begin{eqnarray}
  |\Psi^6_c\rangle &=&\omega_4(| \Phi^{S_{12} =0}_{12} \rangle \otimes
 |\Phi^{S_{34} =0}_{34} \rangle \otimes
 |\Phi^{S_{65} =0}_{65}\rangle) \nonumber \\
&& + \omega_4^2  (|\Phi^{S_{14} =0}_{14} \rangle \otimes
 |\Phi^{S_{25} =0}_{25}\rangle \otimes |\Phi^{S_{36} =0}_{36}\rangle)   \nonumber \\
&&+ \omega_4^3 (|\Phi^{S_{61} =0}_{61}\rangle \otimes
 |\Phi^{S_{23} =0}_{23} \rangle \otimes
 |\Phi^{S_{45} =0}_{45}\rangle) ,
\label{rsol4}
\end{eqnarray}
and its complex conjugate
 $ |\Psi^{6*}_c\rangle$.
 The state in Eq. (\ref{rsol4}) represents 
 the superposition of the three 
%Rumer
 diagrams in Fig. \ref{six}(a) with the 
second term in the sum representing crossing singlets.

\subsection{Producing isotropy in homogeneous states}
Next, we produce isotropy in a given homogeneous state.
 Such states should also  yield  the  entanglement value $E^2_{v, max}$. 
The condition of isotropy in a homogeneous state 
can be expressed  as $\sum_i S_i^{\alpha}|\Psi_N \rangle_{hom}=0$ 
with $\alpha = x, y,$ or $ z$; this  implies that
$\sum_i S_i^+|\Psi_N\rangle_{hom}=0$.
Now, reflection about the z-axis transforms $ \Phi^{S_{ij}=0}_{ij}$ to  
$ -\Phi^{S_{ij}=0}_{ij}$. Then, in a $S_T=0$ state comprising of
$\frac{N}{2}$ singlets, reflection leads to the coefficient  $(-1)^{\frac{N}{2}}$
for the parent state.
Hence,  when isotropy is imposed on the  homogeneous state  of Eq. (\ref{hom}) and 
 all the spins are flipped,  the parent state acquires the coefficient 
$(-1)^{\frac{N}{2}}$.
In fact, in Eq. (\ref{hom}),
the coefficients of the basis state $|\sigma^z_1 \sigma^z_2...\sigma^z_N \rangle$
and its spin flipped version differ only by a factor $(-1)^{\frac{N}{2}}$.
 Thus, the number
of unknown coefficients $e^{i\delta_k}$ in Eq. (\ref{hom}) is reduced to $0.5 \left ({^NC_{\frac{N}{2}}}\right )$.
This also implies that the condition $\sum_i S_i^-|\Psi_N\rangle_{hom}=0$ yields the same equations
as the condition $\sum_i S_i^+|\Psi_N\rangle_{hom}=0$ does. 
Therefore, we can write the four-spin homogeneous
state  as
\begin{eqnarray}
|\Psi_4\rangle_{hom} &=& e^{i \phi_1} [|\uparrow\downarrow\uparrow\downarrow\rangle 
               + |\downarrow\uparrow\downarrow\uparrow\rangle ] \nonumber \\ 
 &+& e^{i\phi_2} [| \uparrow\downarrow\downarrow\uparrow \rangle +
|\downarrow\uparrow\uparrow\downarrow \rangle ] \nonumber \\ 
&+& e^{i\phi_3}[ | \uparrow\uparrow\downarrow\downarrow \rangle +
   | \downarrow\downarrow\uparrow\uparrow \rangle ].
\end{eqnarray}

Furthermore,  in the above four-spin, homogeneous state, the condition of isotropy yields
\begin{eqnarray}
\sum_{i=1}^4 S^+_i |\Psi_4\rangle_{hom} &=&   (e^{i \phi_1}+ e^{i \phi_2}+ e^{i \phi_3})  
     \left [ |\uparrow\uparrow\uparrow\downarrow \rangle 
+ |\uparrow\downarrow\uparrow\uparrow \rangle \right.  \nonumber \\
&& ~~~~~~~~~~~~~~~~~~~~~~~~~~\left . | \uparrow\uparrow\downarrow\uparrow \rangle 
+  |\downarrow\uparrow\uparrow\uparrow \rangle \right]  \nonumber \\
&=&0.
\end{eqnarray}
This leads to the expression $e^{i \phi_1}+ e^{i \phi_2}+ e^{i \phi_3} =0$, i.e., zero-valued
coefficients for all the four basis states. 
There are only two independent solutions to this equation:
% (up to a phase factor) 
$\{e^{i\phi_1} =1,e^{i\phi_2}=\omega_3, e^{i\phi_3} =\omega_3^2 \}$ 
and its complex conjugate. The states 
obtained from these solutions are the same as those  obtained earlier
 by imposing homogeneity on isotropic states,
% and represent  the four-spin maximally entangled state,
i.e.,  $|\Psi^4\rangle$ in Eq. (\ref{psi_hs})
 and its complex conjugate. 

Next, we consider the case of a six-spin, 
homogeneous state and impose isotropy.
We then generate the same entangled states 
$ |\Psi^6_a\rangle$,
$ |\Psi^{6*}_a\rangle$,
$ |\Psi^6_b\rangle$, $ |\Psi^6_c\rangle$, and
  $ |\Psi^{6*}_c\rangle$
 obtained 
in the previous section (see Ref. \cite{suppl} for details).
 Lastly, for any of the states 
$ |\Psi^6_a\rangle$,
$ |\Psi^{6*}_a\rangle$,
$ |\Psi^6_b\rangle$, $ |\Psi^6_c\rangle$, or
  $ |\Psi^{6*}_c\rangle$, we 
observe that the  von Neumann entropy $E^2_v$
has the value 1.921928 which is the
same as that given by $E^2_{v,max}$  in Eq. (\ref{E_vN}).

\subsection{Discussion of the general case}
We will now discuss generating $E^2_{v,max}$ RVB states for the general case of N-spin system.
The number of coefficients $r_i e^{i \theta_i}$,  needed to generate a superposed state $|\Psi^N \rangle $
using all the non-crossing Rumer states, is the same as the 
total  number of linearly-independent Rumer states, i.e.,  
${^N}C_{\frac{N}{2}}-{^N}C_{\frac{N}{2} - 1}$. The number of unknown coefficients $e^{i \delta_k}$
(for the basis states $|\psi^z_k \rangle$) in Eq. (\ref{hom}) is $0.5\left ( {^N}C_{\frac{N}{2}}\right )$.
On setting $|\Psi^N \rangle  = |\Psi_N \rangle_{hom} $ and equating the coefficients
of the various basis states $|\psi^z_k \rangle$, we get
$0.5\left ( {^N}C_{\frac{N}{2}}\right )$  equations; from these equations, on eliminating 
$r_i e^{i \theta_i}$ in terms of the various $e^{i\delta_k}$, we get  
$-0.5\left ( {^N}C_{\frac{N}{2}}\right )+{^N}C_{\frac{N}{2} - 1}$ number of linearly-independent
equations 
in terms of $0.5\left ( {^N}C_{\frac{N}{2}}\right )$  number of unknown coefficients $e^{i\delta_k}$.
Thus, we expect the number of independent $E^2_{v,max}$
RVB states to be equal to  
 the number of unknown $e^{i\delta_k}$ minus the number of independent equations, i.e.,
${^N}C_{\frac{N}{2}}-{^N}C_{\frac{N}{2} - 1}$
which is the  number of linearly-independent Rumer states.
For instance, for $N=8$, we get 21 linearly-independent equations in terms of
35 different $e^{i\delta_k}$; hence, the number of independent $E^2_{v,max}$ RVB states  
is 14 ($=35-21$) which is the total number of non-crossing Rumer states.
We also would like to point out that explicit construction of  $E^2_{v,max}$ RVB states
becomes more and more cumbersome as the spin-system size N increases.

\section{Generating  highly entangled ground states using IRHM}
In this section, we will 
demonstrate  that the $E_{v, max}^2$ entangled states   
obtained earlier, from the homogenization 
of isotropic states or from imposing isotropy on homogeneous states, are the 
ground states of a spin Hamiltonian.  
To this end, we begin with the IRHM Hamiltonian  
\begin{eqnarray}
\!\!\!\!\!\!\!\! H_{\rm IRHM} = J\! \sum_{i,j>i} \!\! \vec{S_i}.\vec{S_j} 
= \frac{J}{2} \! \left [ \!
 \left ( \sum_{i} \vec{S_i} \right )^2 \!\!
 - \! \left ( \sum_{i} \vec{S_i}^2 \right )
 \!\right ] ,
\label{H_gen}
\end{eqnarray} 
and 
show that certain superpositions of the ground states of IRHM  will produce 
the same amount of entanglement as given by Eq. (\ref{E_vN}).
We observe that $[S^z_{Total},H_{\rm IRHM}]=0 $ and  that $[S^2_{Total}, H_{\rm IRHM}]=0$.
In Eq. (\ref{H_gen}), we take
 $J = J^{\star}/(N-1)$ 
(where $ J^{\star}$ is a finite
quantity) so that the energy per site remains finite as $N \rightarrow \infty$.
The eigenstates of $H_{\rm IRHM}$
 correspond to eigenenergies
\begin{eqnarray}
 E_{S_T} = \frac{J}{2} \left [ S_T (S_T + 1) - \frac{3N}{4} \right ], 
 \label{eigenvalue}
\end{eqnarray}
where $S_T$ is the eigenvalue of the total spin.
{From Eqs. (\ref{H_gen}),  
for a homogenized $S_T=0$ state, we get
\begin{eqnarray}
E_{S_T} = J \sum_{i,j>i} \!\! \langle\vec{S_i}.\vec{S_j}\rangle = \frac{3JN(N-1)}{2} \langle{S^z_i}{S^z_j}\rangle ,
 \label{avgenergy}
\end{eqnarray}
which establishes the connection between $E_{S_T}$ and Eq.  (\ref{E_vN}) through Eq. (\ref{Eij}).
Now, any VB state is an eigenstate of $H_{\rm IRHM}$ \cite{suppl}. 
Since, Rumer states are also VB states,  a homogenized linear combination of non-crossing Rumer  states
is also an eigenstate of IRHM with entanglement given by Eq. (\ref{E_vN}).}

{Here, we should mention that the well-known Lipkin-Meshkov-Glick model 
(for a certain choice of parameters) \cite{lmg} is a special case of the IRHM. The IRHM  
with two spins and four spins 
(with spins at the corners of a regular tetrahedron) can be realized from a Hubbard model. 
It has been shown that a zigzag graphene nanodisc can be 
described well by a long-range ferromagnetic Heisenberg model \cite{ezawa}.
 Moreover, the fully connected network  
%which is characterized by a distance-independent hopping 
(which can be mapped onto a spin system with distance-independent spin-spin 
interaction) is a well-studied 
model in the context of excitation energy transfer in light-harvesting 
complexes \cite{fcn}.}

\section{Conclusions.}
In a VB state, both the spins of any two-spin singlet 
{\nolinebreak ( $|\uparrow\downarrow\rangle-|\downarrow\uparrow\rangle$ )}
 are completely unentangled with the rest of the system.
%are
%monogamous (i.e., they cannot be entangled with any other spin),
However, by using a homogenized superposition of VB
states,
% (each of which is a product of $N/2$
%two-spin singlets),
 we managed to distribute entanglement
 efficiently 
such that any pair is maximally entangled with the rest of the RVB system
while concomitantly the constituent spins of the pair are completely unentangled
with each other.  
Now, we know from  Lieb-Mattis theorem \cite{lieb} that 
 states with total spin zero are quite commonly ground states of interacting spin systems.
However, it has not been recognized that one can 
generate high bipartite $E^2_v$ entanglement from such states.
 Our RVB states with maximal $E^2_v$ can be realized physically
in systems such as the
infinite-range, large $U/t$ Hubbard model and infinite-range, hard-core-boson model
 with frustrated hopping \cite{ashvin} when they are at 
half-filling.

\section{Acknowledgments}
One of the authors (S. Y.) would like to thank 
G. Baskaran, R. Simon, and S. Ghosh for valuable discussions.

\clearpage
\newpage
\setcounter{equation}{0}
\setcounter{figure}{0}
\setcounter{table}{0}
\renewcommand{\theequation}{S\arabic{equation}}
\renewcommand{\thefigure}{S\arabic{figure}}
\renewcommand{\thetable}{S\arabic{table}}
\noindent {{\bf Supplemental Material for
``Study of two-spin entanglement in singlet states''}}
\section{ Singlet as a superposition of valence-bond states} 
We will sketch an argument showing that any spin-singlet state can be expressed
 as a superposition of valence-bond (VB) states. 

We first note that there are  $(N-1)!!$ \Big($=(N!)/\big[2^{N/2}(N/2)!\big]$\Big) VB states
each of which is a product of $N/2$ 
dimer states of the form
{\nolinebreak$|\uparrow\downarrow\rangle-|\downarrow\uparrow\rangle$}
 (with no pair of dimers sharing a spin); thus, each VB state has total spin eigenvalue $S_T=0$. 
 It is worth noting that
\begin{eqnarray}
|\Phi^{S_{13} =0}_{13}\rangle \otimes |\Phi^{S_{24} =0}_{24}\rangle &=&
 |\Phi^{S_{12} =0}_{12}\rangle \otimes |\Phi^{S_{34} =0}_{34}\rangle
\nonumber \\
&& +
 |\Phi^{S_{14} =0}_{14}\rangle \otimes |\Phi^{S_{23} =0}_{23}\rangle  ,
\label{lin_dep}
\end{eqnarray}
where   $|\Phi^{S_{ij} =0}_{ij}\rangle \equiv  [ |\uparrow \rangle_i|\downarrow\rangle_j
- |\downarrow\rangle_i|\uparrow\rangle_j]  $ is a dimer for spins
at sites $i$ and $j$ with $S_{ij}$ being the total spin of $S_i$ and $S_j$.
Using the above relation one can establish that there are
only ${^N}C_{\frac{N}{2}}-{^N}C_{\frac{N}{2} - 1}$ linearly-independent
VB states \cite{pauling}.
A particular set of
 linearly-independent VB states are the Rumer states 
\cite{pauling,rumer} that are made up of non-crossing dimers.

Next, we note that a spin eigenfunction with total spin $S_T$ (for a spin system with N spins)
is obtained from the (N-1) electron eigenfunctions by adding or subtracting the spin of the last electron.
Then, the degeneracy $g(N,S_T)$ of the spin $S_T$  state in a N-spin system is given by
\begin{eqnarray}
g(N,S_T)=g(N-1,S_T-\frac{1}{2}) + g(N-1,S_T+\frac{1}{2}) .
\end{eqnarray}
Then,  by induction, it follows that \cite{pauncz}
\begin{eqnarray}
g(N,S_T)= {^N}C_{\frac{N}{2}-S_T}-{^N}C_{\frac{N}{2} -S_T- 1}.
\end{eqnarray}
Thus, the number of linearly-independent spin-singlets (i.e., spin eigenfunctions with $S_T=0$)
is ${^N}C_{\frac{N}{2}}-{^N}C_{\frac{N}{2} - 1}$ which is the same as the number of linearly-independent 
 VB states. Thus, any singlet can be expressed as a linear superposition of VB states.

Of interest are various types of  VB coverings of a lattice. The equal-amplitude superposition of all nearest-neighbor-singlet
VB coverings is a disordered RVB liquid state. When time-reversal symmetry is broken in a valence-bond
spin liquid a chiral spin liquid  is realized. On the other hand, when lattice symmetry is spontaneously
broken, bond-ordered  VB solids result.

\section{ Construction of maximal $E^2_v$ states for six spins} 
As mentioned in the main text, we show two different
ways to construct maximal 
$E_v^2$ states.
\subsection{Generating  homogeneity  in six-spin, isotropic states}
There are five linearly-independent 
states for the case of six spins; they correspond to 
the five non-crossing Rumer diagrams of
 Fig. 3 in the main text.
 Similar to  the four-spin case, we begin with 
the following general linear superposition (mentioned in the main text):
\begin{eqnarray}
|\Psi^6\rangle &=& r_1 e^{i \theta_1} \left[ 
| \Phi^{S_{12} =0}_{12} \rangle \otimes
 |\Phi^{S_{34} =0}_{34} \rangle \otimes
 |\Phi^{S_{56} =0}_{56}\rangle \right]\nonumber \\
&+& r_2 e^{i \theta_2} \left[
|\Phi^{S_{61} =0}_{61}\rangle \otimes
 |\Phi^{S_{23} =0}_{23} \rangle \otimes
 |\Phi^{S_{45} =0}_{45}\rangle \right] \nonumber \\
&+& r_3 e^{i \theta_3} \left[
 |\Phi^{S_{12} =0}_{12}\rangle \otimes |\Phi^{S_{36} =0}_{36} \rangle \otimes |\Phi^{S_{45} =0}_{45}\rangle \right] \nonumber \\
&+& r_4 e^{i \theta_4} \left[
| \Phi^{S_{23} =0}_{23} \rangle \otimes |\Phi^{S_{14} =0}_{14} \rangle \otimes |\Phi^{S_{56} =0}_{56}\rangle \right] \nonumber \\
&+&r_5 e^{i \theta_5} \left[
|\Phi^{S_{16} =0}_{16} \rangle \otimes |\Phi^{S_{25} =0}_{25}\rangle \otimes |\Phi^{S_{34} =0}_{34}\rangle \right], 
\label{sh}
\end{eqnarray}
and obtain solutions that homogenize $|\Psi^6\rangle$, i.e., choose appropriate
coefficients in  $|\Psi^6\rangle$
so as to get  the form  $|\Psi_6\rangle_{hom}$ mentioned in Eq. (11) of the main text. 
Next, we evaluate Eq. (\ref{sh}) and  get the  coefficients 
for the various basis states $|\psi^z_k\rangle$  in Eq. (11) of the main text. The 
expressions for the coefficients $e^{i\delta_k}$ in Eq. (11) of the main text
are shown in Table \ref{table2}. 
Similar to the four-spin case, we 
 make the state in Eq. (\ref{sh}) homogeneous.

\begin{table}[]
\begin{center}
    \begin{tabular}{|l|lllll|}
    \hline
    Basis states $\psi^z_k$  &  \multicolumn{5}{|c|}{Coefficients $e^{i\delta_k}$ in Eq. (11) of main text.} \\  [1ex]\hline
    $\mid \uparrow\downarrow\uparrow\downarrow\uparrow\downarrow \rangle$ & 
     $ r_1 e^{i \theta_1} $ & $- r_2 e^{i \theta_2}$ & $- r_3e^{i\theta_3}$ & $- r_4 e^{i \theta_4}$ & $ - r_5 e^{i \theta_5}$  \\ [1ex]\hline
    $\mid \downarrow\uparrow\uparrow\downarrow\uparrow\downarrow   \rangle $ & 
 $ -r_1 e^{i \theta_1} $ &   & $+r_3e^{i\theta_3}$ & &   \\ [1ex]\hline
    $\mid \uparrow\downarrow\downarrow\uparrow\uparrow\downarrow \rangle $ & 
$ -r_1 e^{i \theta_1} $ &  &  &  & $ + r_5 e^{i \theta_5}$  \\ [1ex]\hline
   $\mid  \downarrow\uparrow\downarrow\uparrow\uparrow\downarrow \rangle $  &
 $ r_1 e^{i \theta_1} $ &  &  & $- r_4 e^{i \theta_4}$ &   \\ [1ex]\hline
    $ \mid \uparrow\downarrow\uparrow\downarrow\downarrow\uparrow \rangle $ & 
$ -r_1 e^{i \theta_1} $ & &  & $+ r_4 e^{i \theta_4}$ &   \\ [1ex]\hline
  $\mid \downarrow\uparrow\uparrow\downarrow\downarrow\uparrow \rangle $ &
$ r_1 e^{i \theta_1} $ &  &  &  & $ - r_5 e^{i \theta_5}$  \\ [1ex]\hline
   $ \mid \uparrow\downarrow\downarrow\uparrow\downarrow\uparrow \rangle$ & 
$ r_1 e^{i \theta_1} $ & & $- r_3e^{i\theta_3}$ &  &   \\ [1ex]\hline
   $ \mid \downarrow\uparrow\downarrow\uparrow\downarrow\uparrow \rangle$ &
 $ -r_1 e^{i \theta_1} $ & $+ r_2 e^{i \theta_2}$ & $+ r_3e^{i\theta_3}$ & $+ r_4 e^{i \theta_4}$ & $ + r_5 e^{i \theta_5}$  \\ [1ex]\hline
 $ \mid \uparrow\uparrow\downarrow\downarrow\uparrow\downarrow \rangle $ & 
   & $+r_2 e^{i \theta_2}$ &  & $+ r_4 e^{i \theta_4}$ &   \\ [1ex]\hline
$ \mid \uparrow\uparrow\downarrow\downarrow\downarrow\uparrow \rangle $ &
  &  &  & $- r_4 e^{i \theta_4}$ &   \\ [1ex]\hline
$ \mid \downarrow\downarrow\uparrow\uparrow\uparrow\downarrow \rangle $ &
&  &  & $+ r_4 e^{i \theta_4}$ &   \\ [1ex]\hline
 $ \mid \downarrow\downarrow\uparrow\uparrow\downarrow\uparrow \rangle $ & 
  & $- r_2 e^{i \theta_2}$ &  & $- r_4 e^{i \theta_4}$ &   \\ [1ex]\hline
 $ \mid \downarrow\downarrow\downarrow\uparrow\uparrow\uparrow \rangle $ & 
  &  & &  & $ - r_5 e^{i \theta_5}$  \\ [1ex]\hline
$ \mid \uparrow\uparrow\uparrow\downarrow\downarrow\downarrow \rangle $ & 
&  & &  & $ + r_5 e^{i \theta_5}$  \\ [1ex]\hline
$ \mid \uparrow\uparrow\downarrow\uparrow\downarrow\downarrow \rangle $ & 
  & $- r_2 e^{i \theta_2}$ &  &  & $ - r_5 e^{i \theta_5}$  \\ [1ex]\hline
$ \mid \downarrow\downarrow\uparrow\downarrow\uparrow\uparrow \rangle $ &
& $+ r_2 e^{i \theta_2}$ &  &  & $ + r_5 e^{i \theta_5}$  \\ [1ex]\hline
$ \mid \downarrow\uparrow\uparrow\uparrow\downarrow\downarrow \rangle $ & 
&  & $- r_3e^{i\theta_3}$ &  &   \\ [1ex]\hline
$ \mid \uparrow\downarrow\downarrow\downarrow\uparrow\uparrow \rangle $ & 
&  & $+ r_3e^{i\theta_3}$ &  &   \\ [1ex]\hline
$ \mid\downarrow\uparrow\downarrow\downarrow\uparrow\uparrow \rangle $ &
 & $- r_2 e^{i \theta_2}$ & $- r_3e^{i\theta_3}$ &  &  \\ [1ex]\hline
$ \mid \uparrow\downarrow\uparrow\uparrow\downarrow\downarrow \rangle $ &
& $+r_2 e^{i \theta_2}$ & $+ r_3e^{i\theta_3}$ &  &  \\ [1ex]\hline
    \end{tabular}
\end{center}
\caption{Basis states $|\psi^z_k\rangle$ of Eq. (11) in the main text for a
six-spin system and the corresponding  coefficients $e^{i\delta_k}$
determined from the equation $|\Psi^6\rangle  =|\Psi_6\rangle_{hom}$.}
\label{table2}
\end{table}
Since there are 
only five linearly-independent Rumer states, we can construct only five linearly-independent 
superpositions of the form of $|\Psi^6\rangle$ given in Eq. (\ref{sh}).
Thus, we can expect that five independent sets of coefficients $r_ie^{i\theta_i}$ 
will produce homogeneity. Here, we present five such linearly-independent solutions:
\begin{eqnarray}
 &1.&\{ r_1=r_2=0; ( r_3 e^{i\theta_3}, r_4 e^{i\theta_4}, r_5 e^{i\theta_5})=(\omega_4, \omega^2_4, \omega_4^3 )  \}, \nonumber \\
 \nonumber \\
&2.&\{ r_1=r_2=0; ( r_3 e^{i\theta_3}, r_4 e^{i\theta_4}, r_5 e^{i\theta_5})
=(\omega^*_4, \omega^{*2}_4, \omega_4^{*3} )  \} ,\nonumber \\
 \nonumber \\
&3.&\{ r_1=r_2=0; ( r_3 e^{i\theta_3}, r_4 e^{i\theta_4}, r_5 e^{i\theta_5})=(-1, +1,-1 )  \}, \nonumber \\
 \nonumber \\
&4.&\{ r_1 e^{i \theta_1} = -1+e^{i \alpha}; r_2 e^{i \theta_2}= 1+e^{i \alpha};\nonumber \\
  &&~~~~~~~~~~~~~~~~~~~~( r_3 e^{i\theta_3}, r_4 e^{i\theta_4}, r_5 e^{i\theta_5})=(-1,-1,-1 )  \}, \nonumber \\
 \nonumber \\
&5.&\{ r_1 e^{i \theta_1} = -1+e^{-i \alpha}; r_2 e^{i \theta_2}= 1+e^{-i \alpha};\nonumber \\
 &&~~~~~~~~~~~~~~~~~~( r_3 e^{i\theta_3}, r_4 e^{i\theta_4}, r_5 e^{i\theta_5})=(-1,-1,-1 )  \} , \nonumber 
\end{eqnarray}
where $\omega_4=e^{i2\pi/4}$ is a fourth root of unity and $e^{i \alpha}$ is a complex number.
In   arriving at the above solutions, 
we used the fact that
when sum of four unit vectors is a zero vector (i.e., when  $\sum_{j=1}^4 e^{i \alpha_j} =0$), 
one pair of the unit vectors will add up to a zero vector with the remaining
pair also producing a zero vector.
In Fig. \ref{sum}, we show one of the three possibilities for $\sum_{j=1}^4 e^{i \alpha_j} =0$. 
Out of the five solutions given above, 
the first and the second solutions produce the states (as mentioned in the main text)
\begin{eqnarray}
 |\Psi^6_a\rangle &=& \omega_4
 (|\Phi^{S_{12} =0}_{12}\rangle \otimes |\Phi^{S_{36} =0}_{36} \rangle \otimes |\Phi^{S_{45} =0}_{45}\rangle) \nonumber \\
&& + \omega_4^2(| \Phi^{S_{23} =0}_{23} \rangle \otimes |\Phi^{S_{14} =0}_{14} \rangle \otimes |\Phi^{S_{56} =0}_{56}\rangle) \nonumber \\
&& + \omega_4^3(|\Phi^{S_{16} =0}_{16} \rangle \otimes |\Phi^{S_{25} =0}_{25}\rangle \otimes |\Phi^{S_{34} =0}_{34}\rangle) ,
\label{sol1}
\end{eqnarray}
and its complex conjugate, respectively, whereas the third solution yields the state
(as given in the main text)
\begin{eqnarray}
 |\Psi^6_b\rangle &=& - (|\Phi^{S_{12} =0}_{12}\rangle \otimes |\Phi^{S_{36} =0}_{36} \rangle \otimes |\Phi^{S_{45} =0}_{45}\rangle) \nonumber \\
&& + (| \Phi^{S_{23} =0}_{23} \rangle \otimes |\Phi^{S_{14} =0}_{14} \rangle \otimes |\Phi^{S_{56} =0}_{56}\rangle) \nonumber \\
&& - (|\Phi^{S_{16} =0}_{16} \rangle \otimes |\Phi^{S_{25} =0}_{25}\rangle \otimes |\Phi^{S_{34} =0}_{34}\rangle).
\label{sol3}
\end{eqnarray}
The states that are obtained from the fourth and the fifth solutions 
can be written as
\begin{eqnarray}
|\Psi_c^6\rangle &=& (-1+e^{i\alpha}) \left[ 
| \Phi^{S_{12} =0}_{12} \rangle \otimes
|\Phi^{S_{34} =0}_{34} \rangle \otimes
|\Phi^{S_{56} =0}_{56}\rangle \right]\nonumber \\
&& + (1+e^{i\alpha}) \left[
|\Phi^{S_{61} =0}_{61}\rangle \otimes
 |\Phi^{S_{23} =0}_{23} \rangle \otimes
 |\Phi^{S_{45} =0}_{45}\rangle \right] \nonumber \\
&&- \left[
 |\Phi^{S_{12} =0}_{12}\rangle \otimes |\Phi^{S_{36} =0}_{36} \rangle \otimes |\Phi^{S_{45} =0}_{45}\rangle \right] \nonumber \\
&&- \left[
| \Phi^{S_{23} =0}_{23} \rangle \otimes |\Phi^{S_{14} =0}_{14} \rangle \otimes |\Phi^{S_{56} =0}_{56}\rangle \right] \nonumber \\
&&- \left[
|\Phi^{S_{16} =0}_{16} \rangle \otimes |\Phi^{S_{25} =0}_{25}\rangle \otimes |\Phi^{S_{34} =0}_{34}\rangle \right] ,
\label{sol4}
\end{eqnarray}
and its complex conjugate, respectively. 
Using $e^{i\alpha}=-i (=-\omega_4)$, the state in  Eq. (\ref{sol4})
can be rewritten as 
\begin{eqnarray}
  |\Psi^6_c\rangle &=&\omega_4(| \Phi^{S_{12} =0}_{12} \rangle \otimes
 |\Phi^{S_{34} =0}_{34} \rangle \otimes
 |\Phi^{S_{65} =0}_{65}\rangle) \nonumber \\
&& + \omega_4^2  (|\Phi^{S_{14} =0}_{14} \rangle \otimes
 |\Phi^{S_{25} =0}_{25}\rangle \otimes |\Phi^{S_{36} =0}_{36}\rangle)   \nonumber \\
&&+ \omega_4^3 (|\Phi^{S_{61} =0}_{61}\rangle \otimes
 |\Phi^{S_{23} =0}_{23} \rangle \otimes
 |\Phi^{S_{45} =0}_{45}\rangle) .
\label{rsol4}
\end{eqnarray}
 The above state represents 
 the superposition of the three 
 diagrams of Fig. 3(a) in the main text.

Evaluating the   tensor products  in the above Eqs. (\ref{sol1}), (\ref{sol3}), and  (\ref{rsol4}),
we rewrite these states
as
 \begin{eqnarray}
\!\!\!\!\!\! |\Psi^6_a\rangle \!\!&=& \!\!
%\frac{1}{\sqrt{20}}
\Big [ \big ( |\uparrow\downarrow\uparrow\downarrow\uparrow\downarrow\rangle 
- |\downarrow\uparrow\downarrow\uparrow\downarrow\uparrow \rangle \big ) 
\nonumber \\
&&
+\omega_4 \big ( |\uparrow\downarrow\uparrow\uparrow\downarrow\downarrow\rangle  
+ |\uparrow\downarrow\downarrow\downarrow\uparrow\uparrow\rangle 
+ |\downarrow\uparrow\uparrow\downarrow\uparrow\downarrow \rangle 
\nonumber \\
&&~~~~~~~
-|\uparrow\downarrow\downarrow\uparrow\downarrow\uparrow\rangle 
-|\downarrow\uparrow\uparrow\uparrow\downarrow\downarrow\rangle 
-|\downarrow\uparrow\downarrow\downarrow\uparrow\uparrow\rangle \big ) 
\nonumber \\
&&
+ \omega_4^2 \big ( |\uparrow\uparrow\downarrow\downarrow\uparrow\downarrow\rangle
+| \uparrow\downarrow\uparrow\downarrow\downarrow\uparrow\rangle
+|\downarrow\downarrow\uparrow\uparrow\uparrow\downarrow\rangle 
\nonumber \\
&&~~~~~~~
-|\uparrow\uparrow\downarrow\downarrow\downarrow\uparrow\rangle
-|\downarrow\uparrow\downarrow\uparrow\uparrow\downarrow\rangle
-|\downarrow\downarrow\uparrow\uparrow\downarrow\uparrow\rangle \big )
\nonumber \\
&&
+\omega_4^3 \big ( |\uparrow\uparrow\uparrow\downarrow\downarrow\downarrow\rangle
+ |\downarrow\downarrow\uparrow\downarrow\uparrow\uparrow\rangle
+ |\uparrow\downarrow\downarrow\uparrow\uparrow\downarrow\rangle 
\nonumber \\
&&~~~~~~~
-|\uparrow\uparrow\downarrow\uparrow\downarrow\downarrow\rangle
-|\downarrow\uparrow\uparrow\downarrow\downarrow\uparrow\rangle 
-|\downarrow\downarrow\downarrow\uparrow\uparrow\uparrow\rangle \big ) \Big ] ,
\label{psi6a}
\end{eqnarray}

\begin{eqnarray}
\!\!\!\!\!\! |\Psi^6_b\rangle \!\!&=& \!\!
%\frac{1}{\sqrt{20}}
\Big [ \big (|\uparrow\downarrow\uparrow\downarrow\uparrow\downarrow\rangle 
- |\downarrow\uparrow\downarrow\uparrow\downarrow\uparrow \rangle \big )
\nonumber \\
&&
 - \big (
  |\uparrow\downarrow\uparrow\uparrow\downarrow\downarrow\rangle  
+ |\uparrow\downarrow\downarrow\downarrow\uparrow\uparrow\rangle 
+ |\downarrow\uparrow\uparrow\downarrow\uparrow\downarrow \rangle
\nonumber \\
&&~~~
 -|\uparrow\downarrow\downarrow\uparrow\downarrow\uparrow\rangle 
-|\downarrow\uparrow\uparrow\uparrow\downarrow\downarrow\rangle 
-|\downarrow\uparrow\downarrow\downarrow\uparrow\uparrow\rangle 
\big )
 \nonumber \\
&&
+ \big (
|\uparrow\uparrow\downarrow\downarrow\uparrow\downarrow\rangle
+| \uparrow\downarrow\uparrow\downarrow\downarrow\uparrow\rangle
+|\downarrow\downarrow\uparrow\uparrow\uparrow\downarrow\rangle 
\nonumber \\
&&~~~
-|\uparrow\uparrow\downarrow\downarrow\downarrow\uparrow\rangle 
-|\downarrow\uparrow\downarrow\uparrow\uparrow\downarrow\rangle
-|\downarrow\downarrow\uparrow\uparrow\downarrow\uparrow\rangle
\big )
 \nonumber \\
&&
- \big (
|\uparrow\uparrow\uparrow\downarrow\downarrow\downarrow\rangle
+ |\downarrow\downarrow\uparrow\downarrow\uparrow\uparrow\rangle
+ |\uparrow\downarrow\downarrow\uparrow\uparrow\downarrow\rangle
\nonumber \\
&&~~~
-|\uparrow\uparrow\downarrow\uparrow\downarrow\downarrow\rangle
-|\downarrow\uparrow\uparrow\downarrow\downarrow\uparrow\rangle 
-|\downarrow\downarrow\downarrow\uparrow\uparrow\uparrow\rangle
\big )
\Big ]  ,
\label{psi6b}
\end{eqnarray} 

and

\begin{eqnarray}
\!\!\!\!\!\! |\Psi^6_c\rangle \!\!&=& \!\!
%\frac{1}{\sqrt{20}}
\Big [
\big (|\uparrow\downarrow\uparrow\downarrow\uparrow\downarrow\rangle 
- |\downarrow\uparrow\downarrow\uparrow\downarrow\uparrow \rangle \big )
\nonumber \\
&&
 +\omega_4 \big (|\uparrow\downarrow\uparrow\downarrow\downarrow\uparrow\rangle  
 + |\uparrow\downarrow\downarrow\uparrow\uparrow\downarrow\rangle
 + |\downarrow\uparrow\uparrow\downarrow\uparrow\downarrow \rangle
 \nonumber \\
&&~~~~~~~
 -|\uparrow\downarrow\downarrow\uparrow\downarrow\uparrow\rangle
 -|\downarrow\uparrow\uparrow\downarrow\downarrow\uparrow\rangle 
 -|\downarrow\uparrow\downarrow\uparrow\uparrow\downarrow\rangle \big )
 \nonumber \\
&&
+ \omega_4^2 \big ( |\uparrow\uparrow\uparrow\downarrow\downarrow\downarrow\rangle 
+ |\uparrow\downarrow\downarrow\downarrow\uparrow\uparrow\rangle
+ |\downarrow\downarrow\uparrow\uparrow\uparrow\downarrow\rangle
\nonumber \\
&&~~~~~~~
-|\uparrow\uparrow\downarrow\downarrow\downarrow\uparrow\rangle
-|\downarrow\uparrow\uparrow\uparrow\downarrow\downarrow\rangle 
-|\downarrow\downarrow\downarrow\uparrow\uparrow\uparrow\rangle \big )
\nonumber \\
&&
+\omega_4^3 \big ( |\uparrow\uparrow\downarrow\downarrow\uparrow\downarrow\rangle
+| \uparrow\downarrow\uparrow\uparrow\downarrow\downarrow\rangle
+|\downarrow\downarrow\uparrow\downarrow\uparrow\uparrow\rangle 
\nonumber \\
&&~~~~~~~
-|\uparrow\uparrow\downarrow\uparrow\downarrow\downarrow\rangle
-|\downarrow\uparrow\downarrow\downarrow\uparrow\uparrow\rangle
-|\downarrow\downarrow\uparrow\uparrow\downarrow\uparrow\rangle
 \big ) \Big ] .
\label{psi6c}
\end{eqnarray}

\subsection{Producing isotropy in six-spin, homogeneous state}
\begin{figure}[]
\begin{center}
\includegraphics[width=2.0in,height=2.0in]{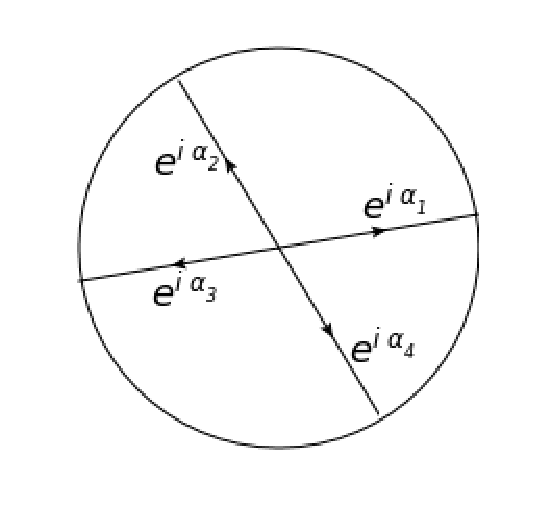}
\caption{Geometric depiction  of zero sum for four unit vectors on
a unit circle in the complex plane. This represents one of the three  possibilities of the zero sum
$\sum_{j=1}^4 e^{i \alpha_j} =0 $. The other two possibilities
are $e^{i\alpha_1} + e^{i\alpha_2} = 0$, $e^{i\alpha_3} + e^{i\alpha_4} = 0$; and 
$e^{i\alpha_1} + e^{i\alpha_4} = 0$, $e^{i\alpha_2} + e^{i\alpha_3} = 0$.}
\label{sum}
\end{center}
\end{figure}

A homogeneous state with six spins can be written as follows:
\begin{eqnarray}
\!\!\!\!\!\!\!
 |\Psi_6\rangle_{hom} = && 
 e^{i \phi_1} \left[ | \uparrow\downarrow\uparrow\downarrow\uparrow\downarrow \rangle -  
   |\downarrow\uparrow\downarrow\uparrow\downarrow\uparrow \rangle \right]    \nonumber \\ 
&+& e^{i \phi_2} \left[  |\downarrow\uparrow\uparrow\downarrow\uparrow\downarrow   \rangle  
- |\uparrow\downarrow\downarrow\uparrow\downarrow\uparrow \rangle  \right]  \nonumber \\ 
&+& e^{i \phi_3} \left[ |\uparrow\downarrow\downarrow\uparrow\uparrow\downarrow \rangle  
- | \downarrow\uparrow\uparrow\downarrow\downarrow\uparrow \rangle \right] \nonumber \\ 
&+& e^{i \phi_4} \left[ | \downarrow\uparrow\downarrow\uparrow\uparrow\downarrow \rangle 
-|\uparrow\downarrow\uparrow\downarrow\downarrow\uparrow \rangle \right]  \nonumber \\ 
&+& e^{i \phi_5} \left[ | \uparrow\uparrow\downarrow\downarrow\uparrow\downarrow \rangle  
-|\downarrow\downarrow\uparrow\uparrow\downarrow\uparrow \rangle \right]  \nonumber \\ 
&+& e^{i \phi_6}  \left[ |\uparrow\uparrow\downarrow\downarrow\downarrow\uparrow \rangle
-|\downarrow\downarrow\uparrow\uparrow\uparrow\downarrow \rangle \right]  \nonumber \\
&+&e^{i \phi_7} \left[  |\downarrow\downarrow\downarrow\uparrow\uparrow\uparrow \rangle 
-|\uparrow\uparrow\uparrow\downarrow\downarrow\downarrow \rangle \right]       \nonumber \\ 
&+& e^{i \phi_8} \left[  |\uparrow\uparrow\downarrow\uparrow\downarrow\downarrow \rangle 
-|\downarrow\downarrow\uparrow\downarrow\uparrow\uparrow \rangle \right] \nonumber \\ 
&+& e^{i \phi_9} \left[ |\downarrow\uparrow\uparrow\uparrow\downarrow\downarrow \rangle 
-|\uparrow\downarrow\downarrow\downarrow\uparrow\uparrow \rangle \right] \nonumber \\
&+&e^{i \phi_{10}} \left[ |\downarrow\uparrow\downarrow\downarrow\uparrow\uparrow \rangle 
-|\uparrow\downarrow\uparrow\uparrow\downarrow\downarrow \rangle \right] .
\label{hom6}
\end{eqnarray}
 For the above  homogeneous state, the isotropy condition 
(expressed as $\sum_{i=1}^6 S^+_i |\Psi_N\rangle_{hom} =0$) 
yields the various basis states and the corresponding  coefficients shown
in Table \ref{table3}. 
\begin{table}[]
\begin{center}
    \begin{tabular}{ | p{3cm} | p{4cm} |}
    \hline
    Basis states contained in $\sum_{i=1}^6 S^+_i | \Psi^6\rangle_{hom}$.   &  
  Zero-valued coefficients of basis states  obtained from using isotropy on $ | \Psi_6\rangle_{hom}$.  \\   \hline
    $\mid \uparrow\uparrow\uparrow\downarrow\uparrow\downarrow \rangle$  & 
         $ e^{i \phi_1} + e^{i \phi_2}+ e^{i\phi_5}  -e^{i \phi_7}$  \\ [1ex] %[1ex] increases linespacing
 \hline

    $\mid \uparrow\downarrow\uparrow\uparrow\uparrow\downarrow   \rangle $ & 
 $ e^{i \phi_1} + e^{i \phi_3}-e^{i\phi_6} -e^{i \phi_{10}} $ \\ [1ex]
\hline

    $\mid \uparrow\downarrow\uparrow\downarrow\uparrow\uparrow \rangle $ & 
$ e^{i \phi_1} - e^{i \phi_4} - e^{i\phi_8} - e^{i \phi_9} $ \\ [1ex]\hline

   $\mid  \downarrow\uparrow\uparrow\uparrow\uparrow \downarrow\rangle $  &
 $e^{i \phi_2} + e^{i \phi_4}+ e^{i\phi_9} -e^{i\phi_6} $ \\ [1ex]\hline

$\mid  \downarrow\uparrow\uparrow\downarrow\uparrow\uparrow \rangle $  &
 $e^{i \phi_2} - e^{i\phi_3} - e^{i\phi_8}  + e^{i \phi_{10}}$  \\[1ex] \hline
 
    $ \mid \uparrow\uparrow\downarrow\uparrow\uparrow\downarrow \rangle $ & 
$ e^{i \phi_3} + e^{i \phi_4}+ e^{i\phi_5} + e^{i \phi_8}$ \\ [1ex]\hline

  $\mid \downarrow\uparrow\downarrow\uparrow\uparrow\uparrow \rangle $ &
$ - e^{i\phi_1}+ e^{i \phi_4} + e^{i \phi_7}+ e^{i\phi_{10}}  $ \\ [1ex]\hline

   $ \mid \uparrow\uparrow\downarrow\downarrow\uparrow\uparrow \rangle$ & 
$ e^{i \phi_5} + e^{i \phi_6} - e^{i\phi_9} + e^{i\phi_{10}} $ \\ [1ex] \hline

   $ \mid \uparrow\uparrow\uparrow\downarrow\downarrow\uparrow \rangle$ &
 $ - e^{i\phi_3}- e^{i\phi_4} + e^{i \phi_6} - e^{i \phi_7}$\\  [1ex]\hline

 $ \mid \uparrow\uparrow\downarrow\uparrow\downarrow\uparrow \rangle $ & 
 $-e^{i\phi_1} - e^{i\phi_2}+ e^{i \phi_6} + e^{i \phi_8}$\\ [1ex]\hline

$ \mid \uparrow\downarrow\downarrow\uparrow\uparrow\uparrow \rangle $ &
 $ -e^{i\phi_2} +e^{i \phi_3} + e^{i \phi_7}  - e^{i\phi_9} $\\ [1ex] \hline

$ \mid \downarrow\downarrow\uparrow\uparrow\uparrow\uparrow \rangle $ &
 $ -e^{i\phi_5} - e^{i \phi_6}  +e^{i \phi_7}   - e^{i\phi_8}$\\ [1ex] \hline

 $ \mid \uparrow\uparrow\uparrow\uparrow\downarrow\downarrow \rangle $ & 
 $- e^{i\phi_7} + e^{i \phi_8}+ e^{i\phi_9}  -e^{i \phi_{10}} $\\ [1ex]\hline

 $ \mid \downarrow\uparrow\uparrow\uparrow\downarrow\uparrow \rangle $ & 
 $- e^{i\phi_1} - e^{i\phi_3} - e^{i\phi_5}  + e^{i \phi_9} $ \\ [1ex] \hline

$ \mid \uparrow\downarrow\uparrow\uparrow\downarrow\uparrow \rangle $ & 
$ - e^{i\phi_2}-e^{i\phi_4} - e^{i\phi_5} - e^{i\phi_{10}}$\\ [1ex] \hline
    \end{tabular}
\end{center}
\caption{The basis states and the corresponding  zero-valued coefficients 
obtained from the condition of isotropy 
$\sum_{i=1}^6 S^+_i|\Psi_6\rangle_{hom}=0$.}
\label{table3}
\end{table}
The set of fifteen equations (corresponding to the zero-valued coefficients of the basis states)
in Table \ref{table3} can be reduced to the following  set of five linearly-independent equations:
\begin{eqnarray}
\label{FLE1}
e^{i \phi_1} + e^{i \phi_2}+ e^{i\phi_5} - e^{i \phi_7} =0  , \\
\label{FLE2}
e^{i \phi_1} + e^{i \phi_3}- e^{i\phi_6} - e^{i \phi_{10}} =0  , \\
\label{FLE3}
e^{i \phi_1} - e^{i \phi_4}- e^{i\phi_8} - e^{i \phi_9} =0  , \\
\label{FLE4}
 e^{i \phi_1} - e^{i \phi_4} - e^{i \phi_7}- e^{i\phi_{10}}  =0  ,\\
\label{FLE5}
 e^{i\phi_1} + e^{i \phi_2} - e^{i \phi_6} - e^{i \phi_8} =0.
\end{eqnarray}
Each of the above equations (\ref{FLE1})--(\ref{FLE5}) can be viewed as the 
sum of four unit vectors on a circle in the complex plane. Since the sum is of the 
form $\sum_{j=1}^4 e^{i \alpha_j} =0 $, it implies 
the conditions shown in Fig. \ref{sum}.
A set of five linearly-independent solutions of 
 Eqs. 
(\ref{FLE1})-(\ref{FLE5})
is given  in Table \ref{table4}.
\begin{table}[]
\begin{center}
    \begin{tabular}{|l|l|l|l|l|l|}
    \hline
     Coefficients in  & \multicolumn{5}{|c|}{Five linearly independent solutions} \\ \cline{2-6}
Eqs. (\ref{FLE1})-(\ref{FLE5})   &Sol. 1 & Sol. 2 &Sol. 3 & Sol. 4 & Sol. 5 \\ \cline{1-6}
    $e^{i\phi_1}$ &1 & 1  & 1 & 1  &  1  \\ [1ex]\hline
   $ e^{i\phi_2} $ &$\omega_4 $  & $\omega^{*}_4$ & -1 &  $-e^{i\alpha}$  &  $-e^{-i\alpha}$  \\ [1ex] \hline
   $e^{i\phi_3}$  & $ \omega_4^3$  & $\omega^{* 3}_4$ & -1 &  $-e^{i\alpha}$ &  $-e^{-i\alpha}$  \\  [1ex] \hline   
   $ e^{i\phi_4}$ & $ -\omega_4^2$  & $ -\omega_4^{* 2}$ & -1 & $e^{i\alpha}$  & $e^{-i\alpha}$   \\  [1ex] \hline   
   $e^{i\phi_5}$ & $\omega_4^2$  &$\omega_4^{* 2}$  & 1 & $e^{i\alpha}$  &  $e^{-i\alpha}$  \\  [1ex] \hline   
   $e^{i\phi_6}$  & $-\omega^2_4$  & $-\omega^{* 2}_4$  & -1 & 1  & 1   \\  [1ex] \hline   
   $e^{i\phi_7}$  & $-\omega_4^3$  & $-\omega_4^{* 3}$ & 1 & 1  &  1  \\  [1ex] \hline   
   $e^{i\phi_8}$ & $ -\omega_4^3$  &  $ -\omega_4^{* 3}$ &1  & $-e^{i\alpha}$  & $-e^{-i\alpha}$    \\  [1ex] \hline   
   $e^{i\phi_9}$ & $ -\omega_4$   &  $ -\omega_4^*$ &1  &  1 & 1    \\  [1ex] \hline   
   $e^{i\phi_{10}}$ &  $ -\omega_4$  & $ -\omega_4^*$  & 1 &    $-e^{i\alpha}$ & $-e^{-i\alpha}$ \\  [1ex] \hline   
 \end{tabular}
\end{center}
\caption{Set of five linearly-independent solutions of  Eqs. (\ref{FLE1})-(\ref{FLE5}).}
\label{table4}
\end{table}
This set of solutions produces the entangled states 
$ |\Psi^6_a\rangle$,
$ |\Psi^{6*}_a\rangle$,
$ |\Psi^6_b\rangle$, $ |\Psi^6_c\rangle$, and
  $ |\Psi^{6*}_c\rangle$ which are the same
as those obtained 
in the main text.

\section{Resonating-valence-bond picture}
Our maximal $E^2_v$ states, that form the  ground states of IRHM, are a new class of RVB states made of homogenized
superposition of VB states.
We will now compare the entanglement properties of our RVB
states and the general RVB states $|\Psi\rangle_{\rm rvb}$ of Ref. \cite{sen}
given below:
\begin{eqnarray}
\!\!\!\!\!\!\!\!\!\!\! |\Psi\rangle_{\rm rvb} =
\sum_{i_{\alpha}\in A;~ j_{\beta}\in B}
f(i_1,...,i_M,j_1,...,j_M)
|(i_1,j_1)...(i_M,j_M)\rangle ,
%\nonumber
\label{rvb}
\end{eqnarray}
where $M$ represents the number of sites in each sub-lattice and $f$ is assumed to be isotropic
 over the lattice. 
Also, $ |(i_k,j_k)\rangle \equiv 
\frac{1}{\sqrt2}(|\uparrow\rangle_{i_k} |\downarrow\rangle_{j_k}
 -|\downarrow\rangle_{i_k} |\uparrow\rangle_{j_k})$ 
denotes the singlet dimer connecting a site in sub-lattice $A$ with another site in sub-lattice $B$.
 The valence-bond basis states (used for the above RVB state $|\Psi\rangle_{\rm rvb}$) form an overcomplete set,  
whereas 
our RVB states are constructed from a complete set of ${^N}C_{\frac{N}{2}}-{^N}C_{\frac{N}{2} - 1}$ states.

The rotational invariance of the two-spin reduced density matrix of a RVB state 
allows us to write it in the form of a Werner state \cite{sen}:
\begin{eqnarray}
\rho_{w}(p) =p|(ij)\rangle \langle (ij)|+\frac{1-p}{4}I_4 ,
\end{eqnarray}
where, for $1/3 < p \leq 1$, the Werner state has the spins at site $i$ and site $j$ entangled with each other.
In  Ref. \cite{sen}, an interesting analysis was carried out
for the examples of RVB gas and RVB liquid.
For the case of the RVB gas, 
the $|\Psi\rangle_{\rm rvb}$ state in Eq. (\ref{rvb}) has $f$ as a constant (corresponding to
equal-amplitude superposition of all bipartite valence-bond coverings).
Consequently, based on the values of $p$, it was concluded 
that  finite-size systems
have a non-zero tangle (or entanglement) between the two sites \cite{sen}.
 Next, for the RVB liquid case
involving equal-amplitude superposition of all  nearest-neighbor-singlet
 valence-bond coverings of a lattice, Monte Carlo calculations 
for a $4 \times 4$ lattice reveals 
 zero (non-zero) tangle
between the two sites for open (periodic) boundary conditions \cite{sen}.
In contrast to this, our  RVB states (with maximal $E^2_v$) yield zero entanglement between the two spins
for all system sizes
(as shown in Sec. II of the main text and below).
It has been demonstrated that the ${\rm SU(2)}$ symmetry of the RVB states
ensures that the two-spin correlation function and the parameter $p$ of the Werner state
are related as \cite{sen}
\begin{eqnarray}
\langle\Psi|\vec{S_i}.\vec{S_j}|\Psi\rangle = -\frac{3}{4}p .
\end{eqnarray} 
Then, since our RVB states produce 
\begin{eqnarray}
\langle S^x_iS^x_j \rangle = \langle S^y_iS^y_j \rangle = \langle S^z_iS^z_j \rangle = \frac{-1}{4(N-1)} ,
\label{SS}
\end{eqnarray}
 it follows that $p=\frac{1}{N-1}$; 
 thus, for systems with even number of spins ($N$), we get zero entanglement between the two sites when $N \ge 4$.
Therefore, we see that our 
RVB states  (among the various RVB states), 
while producing maximum entanglement between a pair and the rest of the system,
yield zero entanglement between the spins of that pair.

\section{Eigenstates of IRHM}
Here, we will show that any VB is an eigenstate of the
 IRHM Hamiltonian (mentioned in the main text)  
\begin{eqnarray}
\!\!\!\!\!\!\!\! H_{\rm IRHM} = J\! \sum_{i,j>i} \!\! \vec{S_i}.\vec{S_j} 
= \frac{J}{2} \! \left [ \!
 \left ( \sum_{i} \vec{S_i} \right )^2 \!\!
 - \! \left ( \sum_{i} \vec{S_i}^2 \right )
 \!\right ] .
\label{H_irhm}
\end{eqnarray} 
The eigenenergies of $H_{\rm IRHM}$
are given by
\begin{eqnarray}
 E_{S_T} = \frac{J}{2} \left [ S_T (S_T + 1) - \frac{3N}{4} \right ], 
 \label{e_value}
\end{eqnarray}
where $S_T$ is the total spin eigenvalue. 
The ground state corresponds to $S_T=0$ which is rotationally
invariant and also implies that $S^z_{T}=0$.
Next, we note the interesting fact  that 
\begin{eqnarray}
[\vec{S_1}.\vec{S_3}+\vec{S_2}.\vec{S_4}+\vec{S_1}.\vec{S_4}+\vec{S_2}.\vec{S_3}] 
|\Phi^{S_{12} =0}_{12}\rangle \otimes |\Phi^{S_{34} =0}_{34}\rangle = 0 .
\label{2dimer}
\end{eqnarray}
%\textcolor{red}
{Since $|\Phi^{S_{12} =0}_{12}\rangle$ is an eigenstate of
$\vec{S_1}.\vec{S_2}$, it is obvious that
 $|\Phi^{S_{12} =0}_{12}\rangle \otimes |\Phi^{S_{34} =0}_{34}\rangle$ 
and the VB containing $|\Phi^{S_{12} =0}_{12}\rangle \otimes |\Phi^{S_{34} =0}_{34}\rangle$
are eigenstates of $\vec{S_1}.\vec{S_2} + \vec{S_3}.\vec{S_4}$. 
 Hence, it is clear that $|\Phi^{S_{12} =0}_{12}\rangle \otimes |\Phi^{S_{34} =0}_{34}\rangle$
is an eigenstate of 
$\sum_{i=1,2,3,4;j>i} \vec{S_i}.\vec{S_j}$. Then, for a system of $2N$ spins,
it follows  by mathematical induction that 
a VB state involving $N$ dimers is an eigenstate of $H_{\rm IRHM}$.}

Alternately, from Eq. (\ref{2dimer}),
we note that  a pair of dimers as well as  a VB containing that pair of dimers
are both eigenstates of  the sum of the 4 inter-dimer interactions corresponding to the two dimers.
For a system of $2N$ spins, since there are ${^N}C_2$ pairs of dimers in the $N$-dimer VB,
there are $4 \times {^N}C_2$ inter-dimer interactions in the $H_{\rm IRHM}$; the VB is an eigenstate
 of the sum of the $4 \times {^N}C_2$ inter-dimer interactions. Furthermore, the VB is also an eigenstate
of the $N$ intra-dimer interactions. Now, there are only ${^{2N}}C_2$ spin interactions (in the
 $H_{\rm IRHM}$) which is equal to the sum of $4 \times {^N}C_2$ (number of inter-dimer interactions) and 
$N$ (number of intra-dimer interactions), i.e., ${^{2N}}C_2= 4 \times {^N}C_2 + N$.
 Thus, the VB is an eigenstate of $H_{\rm IRHM}$.

\end{document}